\documentclass[prb,a4paper,twocolumn,showpacs,floatfix,superscriptaddress,amsmath,amsfonts,amssymb,preprintnumbers,citeautoscript]{revtex4}
\pdfoutput=1
\usepackage[T1]{fontenc}\usepackage[latin1]{inputenc}
\usepackage{dcolumn,graphicx,color,booktabs,microtype,afterpage} \graphicspath{{Figures/}}
\usepackage[autoplay]{animate}
\usepackage[charter,greekuppercase=italicized]{mathdesign}\usepackage{sidecap}
\renewcommand{\figurename}{Fig.}
\renewcommand{\tablename}{Table}
\makeatletter\renewcommand{\fnum@figure}[1]{\figurename~\thefigure~(color online).}\makeatother
\makeatletter\renewcommand{\fnum@table}[1]{\tablename~\thetable.}\makeatother

\newcount\hh \newcount\mm
\hh=\time \divide\hh by 60
\mm=\hh \multiply\mm by 60 \mm=-\mm
\advance\mm by \time
\def\now{\number\hh:\ifnum\mm<10{}0\fi\number\mm}

\usepackage[colorlinks,plainpages=false,linkcolor=blue,urlcolor=blue,citecolor=blue,pdfpagemode=UseNone,pdfstartview=FitBH]{hyperref}

\newcommand{\half}{\frac{1}{\protect\raisebox{0.8pt}{\scriptsize 2}}}
\newcommand{\threehalf}{\frac{3}{\protect\raisebox{0.8pt}{\scriptsize 2}}}

\begin{document}

\makeatletter\renewcommand{\ps@plain}{%
\def\@evenhead{\hfill\itshape\rightmark}%
\def\@oddhead{\itshape\leftmark\hfill}%
\renewcommand{\@evenfoot}{\hfill\small{--~\thepage~--}\hfill}%
\renewcommand{\@oddfoot}{\hfill\small{--~\thepage~--}\hfill}%
}\makeatother\pagestyle{plain}


\title{Superconducting properties and pseudogap from preformed\\ Cooper pairs in the triclinic (CaFe$_{1-x}$Pt$_x$As)$_{10}$Pt$_3$As$_8$}

\author{M.\,A.\,\,Surmach}
\affiliation{Institut f{\"u}r Festk{\"o}rperphysik, TU Dresden, D-01069 Dresden, Germany}

\author{F.\,\,Br{\"u}ckner}
\affiliation{Institut f{\"u}r Festk{\"o}rperphysik, TU Dresden, D-01069 Dresden, Germany}

\author{S.\,\,Kamusella}
\affiliation{Institut f{\"u}r Festk{\"o}rperphysik, TU Dresden, D-01069 Dresden, Germany}

\author{R.\,\,Sarkar}
\affiliation{Institut f{\"u}r Festk{\"o}rperphysik, TU Dresden, D-01069 Dresden, Germany}

\author{P.\,\,Y.\,\,Portnichenko}
\affiliation{Institut f{\"u}r Festk{\"o}rperphysik, TU Dresden, D-01069 Dresden, Germany}

\author{J.\,T.\,\,Park}
\affiliation{Heinz Maier-Leibnitz Zentrum (MLZ), TU M{\"u}nchen, D-85747 Garching, Germany}

\author{G.\,\,Ghambashidze}
\affiliation{Max-Planck-Institut f{\"u}r Festk{\"o}rperforschung, Heisenbergstr.\,1, D-70569 Stuttgart, Germany}

\author{H.~Luetkens}
\affiliation{Laboratory for Muon Spin Spectroscopy, Paul Scherrer Institut, CH-5232 Villigen PSI, Switzerland}

\author{P.~Biswas}
\affiliation{Laboratory for Muon Spin Spectroscopy, Paul Scherrer Institut, CH-5232 Villigen PSI, Switzerland}

\author{W.~J.~Choi}
\affiliation{Department of Emerging Materials Science, DGIST, Daegu 711-873, Republic of Korea}

\author{Y.~I.~Seo}
\affiliation{Department of Emerging Materials Science, DGIST, Daegu 711-873, Republic of Korea}

\author{Y.~S.~Kwon}
\affiliation{Department of Emerging Materials Science, DGIST, Daegu 711-873, Republic of Korea}

\author{H.-H.~Klauss}
\affiliation{Institut f{\"u}r Festk{\"o}rperphysik, TU Dresden, D-01069 Dresden, Germany}

\author{D.~S.\,\,Inosov}\email[Corresponding author: \vspace{8pt}]{Dmytro.Inosov@tu-dresden.de}
\affiliation{Institut f{\"u}r Festk{\"o}rperphysik, TU Dresden, D-01069 Dresden, Germany}

\begin{abstract}
\noindent Using a combination of muon-spin relaxation ($\mu$SR), inelastic neutron scattering (INS) and nuclear magnetic resonance (NMR), we investigated the novel iron-based superconductor with a triclinic crystal structure (CaFe$_{1-x}$Pt$_x$As)$_{10}$Pt$_3$As$_8$ ($T_\text{c}=13$\,K), containing platinum-arsenide intermediary layers. The temperature dependence of the superfluid density obtained from the $\mu$SR relaxation-rate measurements indicates the presence of two superconducting gaps, $\Delta_\text{1}\gg\Delta_\text{2}$. According to our INS measurements, commensurate spin fluctuations are centered at the $(\piup,\, 0)$ wave vector, like in most other iron arsenides. Their intensity remains unchanged across $T_\text{c}$, indicating the absence of a spin resonance typical for many Fe-based superconductors. Instead, we observed a peak in the spin-excitation spectrum around $\hslash\omega_0=7$\,meV at the same wave vector, which persists above $T_\text{c}$ and is characterized by the ratio $\hslash\omega_0/k_\text{B}T_\text{c}\approx6.2$, which is significantly higher than typical values for the magnetic resonant modes in iron pnictides ($\sim$\,4.3). The temperature dependence of magnetic intensity at 7\,meV revealed an anomaly around $T^\ast=45$\,K related to the disappearance of this new mode. A suppression of the spin-lattice relaxation rate, $1/T_1T$, observed by NMR immediately below $T^\ast$ without any notable subsequent anomaly at $T_{\rm c}$, indicates that $T^\ast$ could mark the onset of a pseudogap in (CaFe$_{1-x}$Pt$_x$As)$_{10}$Pt$_3$As$_8$, which is likely associated with the emergence of preformed Cooper pairs.
\end{abstract}

\pacs{74.70.Xa 74.25.Ha 78.70.Nx\vspace{-0.7em}}

\maketitle

\vspace{-5pt}\section{Introduction}\vspace{-5pt}

The recently discovered iron-arsenide superconductors of the novel 10-3-8 structure type containing platinum arsenide intermediary layers, (CaFe$_{1-x}$Pt$_x$As)$_{10}$Pt$_3$As$_8$,
are a rare example of an unconventional superconductor family with a triclinic crystal structure \cite{NiAllred11, Lohnert_Sturzer_2011, Kakiya_Kudo_2011}. Their structural and physical properties still remain a subject of current investigation \cite{Xiang_Luo_2012, Zhou_Koutroulakis_2013, Sturzer_Friederichs_2013, Ni_Straszheim_2013}. An interesting peculiarity of these systems is that they display a large difference in their optimal critical temperature, $T_\text{c}$, and ground-state properties depending on the amount of Pt in the intermediary Pt$_{n}$As$_{8}$ spacer layer ($n=3$ or 4) \cite{NiAllred11, Sturzer_Derondeau_2012, Xiang_Luo_2012}. While the parent compound of the 10-3-8 family is not superconducting and only displays superconductivity upon doping \cite{Xiang_Luo_2012, Ni_Straszheim_2013, Sturzer_Derondeau_2014}, in the related 10-4-8 family the optimal $T_\text{c} \approx 35$\,K is reached already in the stoichiometric parent phase and can only be suppressed by doping \cite{NiAllred11, Sturzer_Derondeau_2012}. Still, the identical $T_\text{c}$ for the undoped 10-4-8 and optimally rare-earth doped 10-3-8 compounds \cite{Sturzer_Derondeau_2014} reveal their close resemblance behind structural differences.

Since Pt, which acts as a dopant, is already present in the intermediary layers, growing a large single crystal of the stoichiometric parent compound without any Pt doping on the Fe site still represents a challenge. A further complication comes from the coexistence of multiple twin domains in the same crystal due to the triclinic lattice structure \cite{Cho_Tanatar_2012}. Initially, there was no consensus about the presence or nature of the antiferromagnetic (AFM) order in the 10-3-8 parent compound. At first, no such order was observed \cite{Xiang_Luo_2012}, unlike in most other iron-pnictide families, whereas more recent results from nuclear magnetic resonance (NMR) \cite{Zhou_Koutroulakis_2013}, muon-spin relaxation ($\mu$SR) \cite{Sturzer_Friederichs_2013}, and transport measurements \cite{Ni_Straszheim_2013} revealed AFM order with a N\'{e}el temperature, $T_\text{N}$, as high as 100\,K. However, a wide AFM transition with an onset around 120\,K, derived from $\mu$SR \cite{Sturzer_Friederichs_2013}, stays in contrast to the much sharper transition at $T_\text{N}=100$\,K observed in NMR data \cite{Zhou_Koutroulakis_2013}, suggesting a possible sample dependence of the magnetic properties. Only the most recent x-ray and neutron diffraction measurements \cite{Sapkota_Tucker_2014} could demonstrate the presence of both a second-order structural transition at $T_\text{s}=110(2)$\,K and a magnetic transition to the long-range stripe-AFM phase at $T_\text{N}=96(2)$\,K on the same single crystal. Still, the somewhat lower value of $T_\text{N}$ in this sample as compared to those reported for powders of the same compound could indicate that the crystal slightly deviates from the stoichiometric composition.

Band structure calculations \cite{Berlijn_2014} and angle-resolved photoemission measurements \cite{ThirupathaiahSturzer13, Neupane_Liu_2012} reported that the Fermi surface of iron-platinum arsenides must be highly two-dimensional and similar to that of the usual Fe-based superconductors, suggestive of magnetic excitations peaked at the $(\pi, 0)$ wave vector. This has been directly confirmed in a recent neutron diffraction experiment \cite{Sapkota_Tucker_2014}, which revealed magnetic Bragg reflections consistent with a stripe-AFM order in the $ab$ plane with a ferromagnetic alignment of spins along the $c$ direction. Moreover, intense and steep magnetic excitations were observed at the $(\piup, 0)$ propagation vector in the undoped superconducting 10-4-8 compound by inelastic neutron scattering (INS) \cite{Sato_Kawamata_2011}, similar to those found in most other pnictides \cite{Johnston_2010, Steward_2011}.

It is well known that in most families of iron arsenides the structural transition occurs at a higher temperature, compared to $T_\text{N}$. The region between the two transitions in the phase diagram is typically associated with the much discussed spin-nematic state, which breaks the fourfold rotational symmetry while preserving the translational symmetry of the electronic subsystem \cite{Fernandes_Chubukov_2014, Davis_Hirschfeld_2014}. Therefore, compounds with a large difference between $T_\text{s}$ and $T_\text{N}$, such as NaFeAs \cite{Johnston_2010, Paglione_Greene_2010, Yi_Lu_2012}, represent perfect candidates for studying this enigmatic electronic phase. In the 10-3-8 parent compound, the difference of 14\,--\,20\,K between $T_\text{N}=100$\,K derived from the NMR measurements \cite{Zhou_Koutroulakis_2013} and $T_\text{s}=120$\,K measured directly by x-ray diffraction \cite{Sturzer_Friederichs_2013}, or between $T_\text{s}=110$\,K and $T_\text{N}=96$\,K obtained on the same single crystal by neutron diffraction \cite{Sapkota_Tucker_2014} is unprecedentedly high among all known iron-pnictide compounds, making this system especially promising for experimental investigations of the nematic state.

Turning back to the superconducting properties, we note that superconductivity in the 10-3-8 compounds can be tuned by substituting platinum \cite{NiAllred11, Lohnert_Sturzer_2011, Xiang_Luo_2012, Sturzer_Friederichs_2013, Cho_Tanatar_2012, Kim_Ronning_2012, Kakiya_Kudo_2011, Sturzer_Derondeau_2012, Nohara_Kakiya_2012, Tamegai_Ding_2013, Watson_McCollan_2014, Gao_Sun_2014} or other transition metals \cite{SturzerKessler14} on the Fe site, by trivalent metal doping (Y, La--Sm, Gd--Lu) on the Ca site \cite{Ni_Straszheim_2013, Sturzer_Derondeau_2012, Kim_Sturzer_2013, Sturzer_Derondeau_2014}, or by applying pressure \cite{Nohara_Kakiya_2012, Gao_Sun_2014}. Upon Pt substitution, 10-3-8 compounds remain not superconducting below 3\% doping level \cite{Sturzer_Derondeau_2012}, which correlates to the phase diagrams of most other FeAs superconductors with transition metal substitution on the Fe site, where doping suppresses the spin-density-wave (SDW) state of the parent compound \cite{Johnston_2010, Steward_2011}. A very similar phase diagram is observed upon the application of hydrostatic pressure to the undoped 10-3-8 compound \cite{Gao_Sun_2014}, where pressure-induced superconductivity appears in the 3.5--7\,GPa pressure range with maximal $T_{\rm c}=8.5$\,K at 4.1\,GPa. The superconducting dome follows immediately after the AFM order is suppressed both in doping and pressure phase diagrams \cite{Cho_Tanatar_2012, Ni_Straszheim_2013, Gao_Sun_2014}. Upon rare-earth doping, superconductivity in the phase diagram appears independently of the type of the dopant element, revealing a universal dependence of $T_{\rm c}$ on concentration with an optimum around $0.13{\rm e^{-}/FeAs}$ \cite{Sturzer_Derondeau_2014}.

To date, several essential properties of the superconducting state in 10-3-8 materials are known. For instance, a large anisotropy of the upper critical fields $H^{\parallel}_\text{c2}$ and $H^{\bot}_\text{c2}$ (parallel and perpendicular to the FeAs planes) has been demonstrated \cite{NiAllred11, Xiang_Luo_2012, Ni_Straszheim_2013, Nohara_Kakiya_2012, Watson_McCollan_2014}, reaching $\gamma_{H} = H^{\parallel}_\text{c2}(T)/ H^{\bot}_\text{c2}(T)\approx10$ upon approaching $T_\text{c}$, which makes this compound one of the most two-dimensional among known Fe-based superconductors \cite{Xiang_Luo_2012, Watson_McCollan_2014}. Low-temperature coherence lengths are estimated to be $\xi_{\parallel}(0)=50$\,\AA\ and $\xi_{\bot}(0)=12$\,\AA\ \cite{NiAllred11}, comparable to other arsenides \cite{Johnston_2010, Steward_2011}. The normalized heat capacity jump, $\Delta C/\gamma_\text{n}T_\text{c}$, for the La-doped 10-3-8 compound was reported to be 0.37 (Ref.\,\citenum{Kim_Sturzer_2013}) or 0.81 (Ref.\,\citenum{Ni_Straszheim_2013}), well below the weak-coupling prediction of 1.43 resulting from the Bardeen-Cooper-Schrieffer (BCS) theory. The in-plane London penetration depth $\lambda_{ab}$ was obtained with a value of approximately 1000~nm \cite{Kim_Ronning_2012}, which is substantially higher than typical values for pnictides \cite{Johnston_2010, Steward_2011}. The existence of multiple superconducting gaps has been suggested from the temperature dependencies of critical fields and the London penetration depth \cite{Cho_Tanatar_2012, Watson_McCollan_2014}, yet no direct spectroscopic measurements of the gap magnitude has so far been obtained, to the best of our knowledge. There are additional indications that the anisotropy of the superconducting gap increases toward the edges of the superconducting dome as compared to the optimal doping \cite{Cho_Tanatar_2012}. While direct measurements of the superconducting order parameter remain scarce, it has already been suggested that iron-platinum arsenides could share the same pairing mechanism with other iron-based superconductors, mediated by the AFM spin fluctuations \cite{Zhou_Koutroulakis_2013, Watson_McCollan_2014}. A verification of this hypothesis would require the knowledge of the spin-fluctuation spectrum and the order-parameter symmetry in these new compounds, which we address in the present work.

\begin{figure}[t]
\includegraphics[width=1\columnwidth]{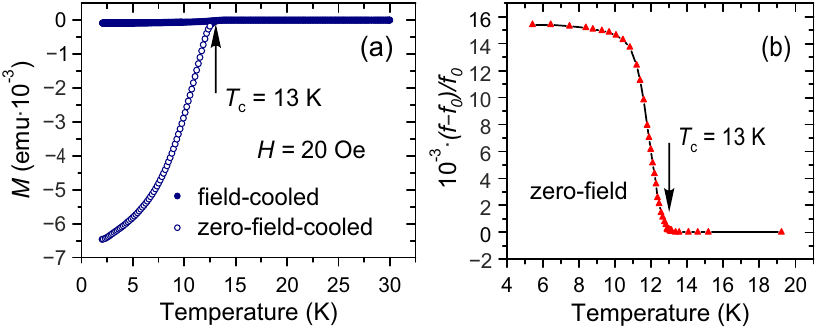}
\caption{(a) Magnetization curve of the optimally doped 10-3-8 compound from the field-cooled and zero-field-cooled measurements. (b) Temperature dependence of the ac susceptibility, quantified by a relative resonant frequency shift, $(f-f_0)/f_0$, of an \emph{in situ} NMR coil on the same sample.}
\label{fig:magnetization}\vspace{-5pt}
\end{figure}

\vspace{-5pt}\section{Experimental results}\vspace{-5pt}

In this paper, we present a combined study of the superconducting properties and spin dynamics in (CaFe$_{1-x}$Pt$_x$As)$_{10}$Pt$_3$As$_8$ \mbox{($x\approx0.06$)} performed by $\mu$SR, INS and NMR measurements on the same single-crystalline sample. The crystal with a mass of $\sim$\,0.5\,g has been grown by the Bridgman method with a molybdenum crucible sealed by the arc welding system\cite{Song_Ghim_2010}. Magnetization measurements presented in Fig.\,\ref{fig:magnetization}\,(a) reveal a sharp superconducting transition at $T_\text{c}=13$\,K, consistent with the optimal doping. This result is also confirmed by an ac susceptibility measurement, performed prior to NMR experiments by tracking the relative resonance frequency shift, $(f-f_0)/f_0$, of the NMR resonance circuit, as shown in Fig.\,\ref{fig:magnetization}\,(b).

\vspace{-5pt}\subsection{$\mu$SR measurements}\vspace{-5pt}

We start the presentation of our results with the $\mu\text{SR}$ measurements, performed at the the Dolly instrument of the Paul Scherrer Institute (Villigen, Switzerland). The single crystal of (CaFe$_{1-x}$Pt$_x$As)$_{10}$Pt$_3$As$_8$ with a mass of $\sim\kern.7pt$0.5\,g was glued to an aluminum foil and fixed to a holder so that crystallographic $ab$ plane was perpendicular both to the direction of the magnetic field and the muon beam. The muon-spin polarization was rotated at 45$^\circ$ with respect to the applied field. The signal was picked up from the left and right pair of detectors. The measurements were done in the temperature range between 2\,K and 25\,K.

First, we performed characterization experiments to exclude the presence of foreign phases. A measurement at $T=2$\,K performed in a weak transverse field of 20\,G showed no fast-depolarizing fraction of the muon asymmetry to within 10\%, thereby excluding any significant amounts of impurity phases with static magnetism (either ordered or disordered) in the sample. As the next step, we verified the absence of any non-superconducting paramagnetic inclusions apart from the main superconducting phase. In the superconducting state, magnetic flux lines penetrate the superconductor in the form of quantized vortices, leading to an inhomogeneous distribution of magnetic field inside the sample, which causes an additional depolarization of the muons. If the sample is cooled down in an applied magnetic field, due to the large value of the London penetration depth \cite{Kim_Ronning_2012}, the well ordered vortex lattice causes only a slow additional depolarization, comparable to the nuclear contribution. To better separate these components, we prepared the sample by cooling it down below $T_{\rm c}$ in zero field and then applying a transverse field of 1000\,G, which resulted in a much more disordered vortex distribution and a much faster depolarization rate. At several measured temperatures, the signal completely depolarized on the time scales $\lesssim$\,2$\mu$s, indicating 100\% superconducting volume fraction with no foreign phases.

\begin{figure}[t]
\includegraphics[width=1\columnwidth]{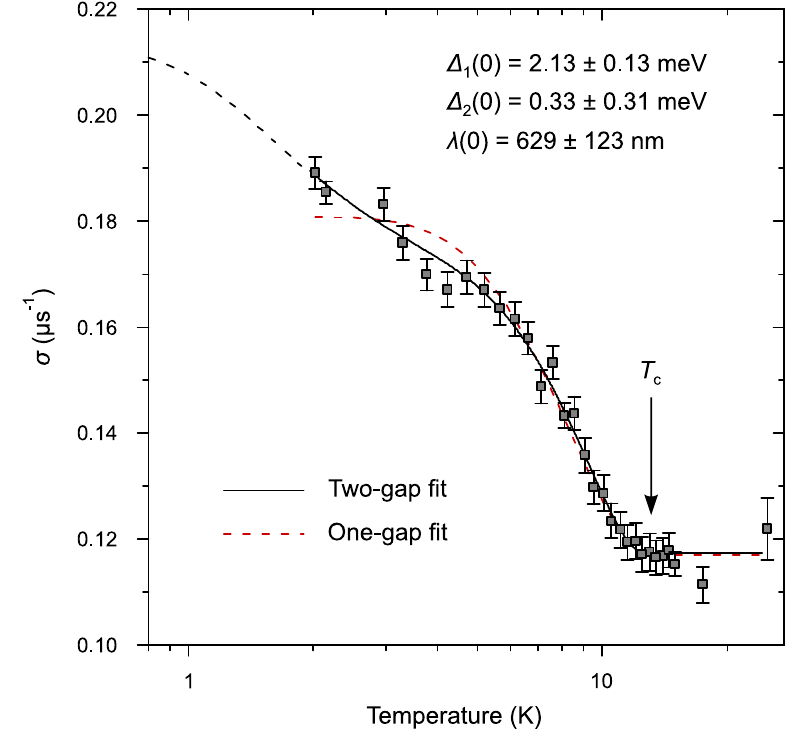}
\caption{Temperature dependence of the muon depolarization rate, $\sigma$, measured in a $\mu$SR experiment and fitted as described in the text (solid and dashed lines).}
\label{fig:muSR}\vspace{-5pt}
\end{figure}

Further, the muon depolarization rate, $\sigma$, has been measured systematically as a function of temperature in a field-cooled experiment. Using the result of our previous measurement, at this stage of data analysis we fixed the superconducting volume fraction to 100\% and fitted the $\mu$SR signal to the following model:
\begin{equation}
A(t)=A\exp(-\sigma^2t^2/2)\cos(\gamma Bt+\phi),
\end{equation}
where $A$ is the initial asymmetry, $\gamma/2\pi= 135.5$~MHz/T is the muon gyromagnetic ratio, and $\phi$ is the initial phase of the muon spins. The resulting temperature dependence of the relaxation rate is presented in Fig.\,\ref{fig:muSR}. It consists of two components:
\begin{equation}
\sigma^2=\sigma_{\rm n}^2+\sigma_{\rm sc}^2,
\label{Eq:SigmaTotal}
\end{equation}
where $\sigma_{\rm n}$ is the constant magnetic nuclear depolarization rate and $\sigma_{\rm sc}$ describes the additional temperature-dependent depolarization due to the vortex lattice in the superconducting state. This latter term is directly proportional to the superfluid density \cite{Brandt88} via
\begin{equation}
\sigma(T)\approx7.09\cdot 10^4\,\text{nm}^2\mu\text{s}^{-1} \cdot\lambda^{-2}(T),
\label{Eq:sigma}
\end{equation}
where the prefactor is given for the high-anisotropy limit \cite{Brandt88, NiedermayerBernhard02}. In turn, $\lambda^{-2}(T)$ contains information about the superconducting order parameter of the compound \cite{PooleProzorov14}.

Following the methodology developed in Refs.~\citenum{Evtushinsky_Inosov_2009, Khasanov_Evtushinsky_2009} for Fe-based superconductors, we describe $\lambda^{-2}(T)$ using the following model for a superconductor with two $s$-wave gaps, $\Delta_1$ and $\Delta_2$:\vspace{-3pt}
\begin{equation}
\frac{1}{\lambda^2(T)}=\lambda^{-2}(0)\left[1 - \alpha M_1(\Delta_1,T) - (1-\alpha)M_2(\Delta_2,T)\right].\hspace{-2pt}\vspace{-1pt}
\label{Eq:LambdaT}
\end{equation}
Here $\lambda^{-2}(0)$ is the superfluid density at $T=0$, $\alpha$ describes the relative contributions of the two gaps, and $M_{1}$ and $M_{2}$ are empirical functions described in Ref.~\citenum{Evtushinsky_Inosov_2009}. The temperature dependence of the superconducting gap, $\Delta(T)$, was approximated by \cite{Gross_Chandrasekhar_1986}\vspace{-6pt}
\begin{equation}
\Delta(T)=\Delta(0)\cdot \tanh \left( \frac{\pi}{2}\cdot\sqrt{\frac{T_\text{c}}{T}-1}\right).
\label{Eq:DeltaT}
\end{equation}
Substituting sequentially Eqs.~(\ref{Eq:DeltaT}), (\ref{Eq:LambdaT}), and (\ref{Eq:sigma}) into Eq.~(\ref{Eq:SigmaTotal}) yields the fitting function $\sigma\left[T, \Delta_1, \Delta_2, \lambda(0), \alpha, \sigma_n\right]$ used to fit the relaxation-rate data in Fig.\,\ref{fig:muSR}, as shown with the solid line.

Parameters $\Delta_{1}$, $\Delta_{2}$, $\lambda(0)$, $\alpha$ and $\sigma_\text{n}$ were fitted simultaneously, while $T_\text{c}$ was kept fixed to the previously determined value. The two-gap model results in a significantly better fit as compared to a single gap (see Fig.\,\ref{fig:muSR}), which agrees with conclusions of an earlier study performed on 10-3-8 samples with a similar composition \cite{Cho_Tanatar_2012}. The resulting gap magnitudes, $\Delta_{1}=2.13\pm0.13$\,meV and $\Delta_{2}=0.33\pm0.31$\,meV, correspond to the $2\Delta/k_\text{B}T_\text{c}$ ratios of 3.8 and 0.58, respectively, which are in good agreement with those of other unconventional two-gap superconductors with similar critical temperatures \cite{Inosov_Park_2011}. For the larger gap, the $2\Delta/k_\text{B}T_\text{c}$ ratio is only slightly higher than the value of 3.53 predicted by the BCS theory, which puts the (CaFe$_{1-x}$Pt$_x$As)$_{10}$Pt$_3$As$_8$ superconductor in the weak-coupling limit. The London penetration depth of $\lambda(0)=629\pm123$\,nm, resulting from our fit, is somewhat higher than typical values for iron pnictides \cite{Johnston_2010}, but nearly twice lower than the value of 1200\,nm reported for a Pt-doped 10-3-8 superconductor with a slightly lower $T_{\rm c}$ of 11\,K from magnetic force microscopy \cite{Kim_Ronning_2012}. It is also more consistent with the empirical Uemura scaling \cite{UemuraLuke89}, generalized for iron-arsenide superconductors in Ref.\,\citenum{BendeleWeyeneth10}.

It has to be noted that the model used here to analyze the $\mu$SR data is unaware of the pseudogap revealed in the INS and NMR experiments, as will be described below. Despite the good description of the experimental data, in the prospect of these results we should consider the large energy gap ($\Delta_{1}$) obtained from the superfluid density just as an effective energy scale related to the superconducting transition, rather than the actual gap in the density of states (DOS). As will be seen from the NMR results, the actual energy gap is both larger in magnitude and opens up at a higher temperature much above $T_{\rm c}$.

\vspace{-5pt}\subsection{INS measurements}\vspace{-5pt}

\begin{figure}[b]
\includegraphics[width=\columnwidth]{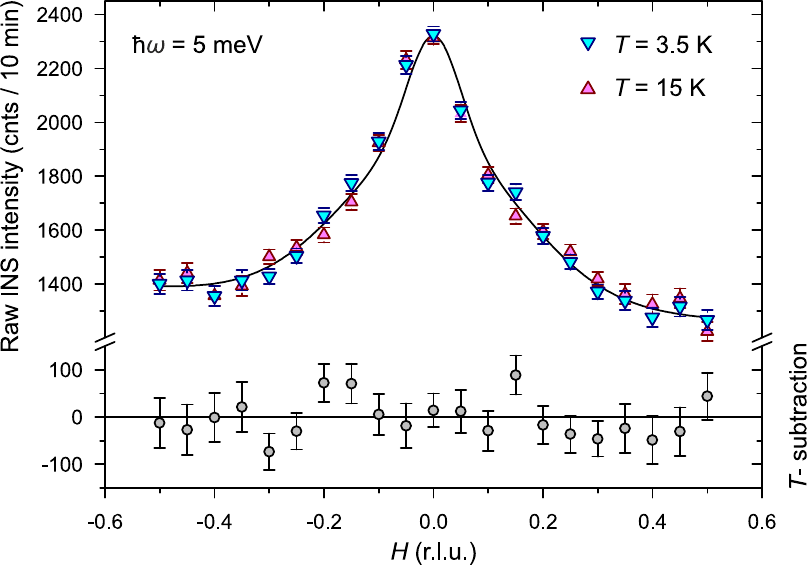}
\caption{Unprocessed INS data, showing transverse scans through $\bigl(0\,\half\bigr)$ at 5\,meV energy transfer for two temperatures: $T=3.5\,\text{K}<T_\text{c}$ ($\triangledown$) and $T=15\,\text{K}>T_\text{c}$ ($\vartriangle$), along with their difference shown below (circles). Solid lines are guides to the eyes.}
\label{fig:transverse}\vspace{-4pt}
\end{figure}

Next, we turn to the presentation of the INS results. For neutron-scattering experiments, we mounted the sample on an aluminum holder and oriented it using neutron diffraction at the full-circle diffractometer Morpheus (PSI, Switzerland) with its $c$-axis pointing upwards, so that the scattering plane is parallel to the FeAs layers. INS measurements were performed using the triple-axis spectrometer PUMA (MLZ, Garching) with a fixed final neutron wave vector of $k_\text{f}=2.662\,\text{\AA}^{-1}\!$.

The low-symmetry triclinic unit cell of the 10-3-8 compounds, containing 10 Fe atoms due to the superstructure of vacancies in the PtAs layer, complicates the description of the reciprocal space in the natural crystallographic notation. Therefore, we will use the unfolded Brillouin zone (BZ) notation, which is usually introduced to simplify the description of the reciprocal space in iron pnictides \cite{ParkInosov10}. It corresponds to the Fe sublattice with one Fe atom per unit cell, whose symmetry in the 10-3-8 systems is very close to tetragonal. Indeed, from the high-temperature lattice parameters of the undoped (CaFeAs)$_{10}$Pt$_3$As$_8$ (Ref.\,\citenum{NiAllred11}), one obtains the following parameters for the Fe sublattice: $a=2.769$\,\AA, $b=2.771$\,\AA, $c=10.272$\,\AA, $\alpha=90.009^{\circ}$, $\beta=90.048^{\circ}$, $\gamma=90.035^{\circ}$. The minor difference between $a$ and $b$ and the deviations of the angles from 90$^\circ$ can be neglected to a good approximation, resulting in a tetragonal unit cell. This fact is best illustrated by the recent x-ray diffraction data on the 10-3-8 parent compound \cite{Sapkota_Tucker_2014}.

\begin{figure}[b!]\vspace{-5pt}
\includegraphics[width=1\columnwidth]{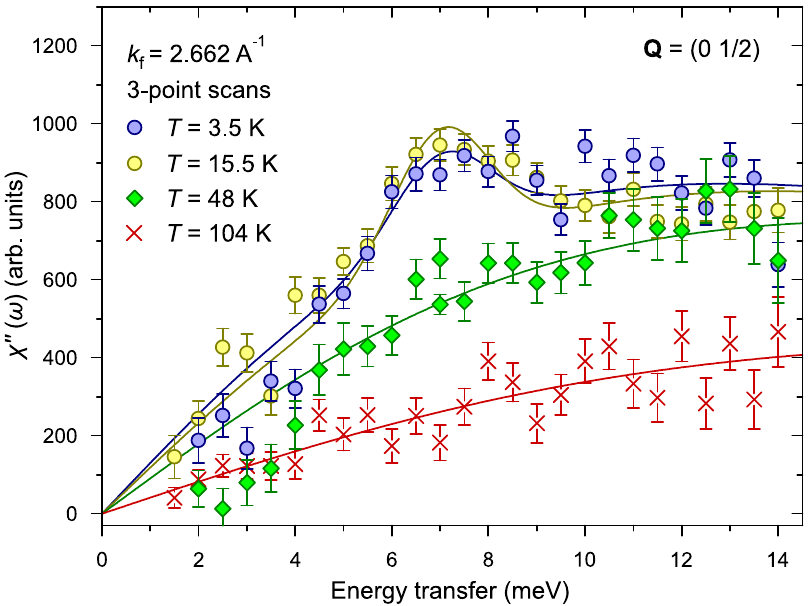}
\caption{Dynamical spin susceptibility, $\chi''(\mathbf{Q}, \omega)$ at $\mathbf{Q}=\bigl(0 \half\bigr)$. The solid lines in high-temperature datasets are fits to the Moriya formula (see text). At low temperatures, an additional Gaussian peak is added to the fitting function to describe the anomalous peak appearing at 7\,meV energy transfer.\vspace{-4pt}}
\label{fig:SQw}
\end{figure}

In Fig.\,\ref{fig:transverse}, we present constant-energy scans through $\bigl(0\,\half\bigr)$ along the transverse direction at two temperatures below and immediately above $T_{\rm c}$: 3.5\,K and 15\,K. A well-centered peak originating from the magnetic signal coincides with the same commensurate $(0,\,\piup)$ wave vector vector as in most other iron-arsenide systems \cite{Johnston_2010, Steward_2011}. However, Fe-based superconductors typically exhibit a magnetic resonant mode \cite{Johnston_2010}, manifested in the transfer of the magnetic spectral weight below the superconducting transition from the low-energy spin-gap region to higher energies immediately under $2\Delta$. The resonance energy scales linearly with $T_{\rm c}$, approximately following a simple empirical rule established for a broad class of Fe-based systems: $\hslash\omega_{\rm res}\approx4.3k_{\rm B}T_{\rm c}$ (Refs.\,\citenum{ParkInosov10, WangLuo10, ShamotoIshikado10}). For our compound with $T_\text{c}=13$\,K, it is therefore expected to appear around 5\,meV, where the scans in Fig.\,\ref{fig:transverse} were measured. Contrary to this expectation, we observe no change in the INS intensity between the datasets acquired below and above $T_{\rm c}$, as best seen in the subtraction of low- and high-temperature data shown at the bottom of Fig.~\ref{fig:transverse}. Hence, the magnetic resonant mode in the conventional sense is absent in (CaFe$_{1-x}$Pt$_x$As)$_{10}$Pt$_3$As$_8$.

To observe the overall shape of the signal and to confirm its magnetic origin, we performed energy scans at the same $\textbf{Q}$ for several temperatures. All scans were complemented by background measurements on both sides of the peak, which were averaged and subtracted from the midpoint intensity to obtain the scattering function, $S(\mathbf{Q},\omega)$. The imaginary part of the dynamical spin susceptibility (Fig.\,\ref{fig:SQw}) was then obtained via the fluctuation-dissipation relation: $\chi''(\mathbf{Q},\omega)=(1-\mathrm{e}^{-\hslash\omega/k_{\rm B}T})\,S(\mathbf{Q},\omega)$. The signal is clearly suppressed towards higher temperatures and its overall shape can be well described by the Moriya formula (solid lines), derived for nearly AFM metals \cite{Moriya}, which was previously shown to hold well for the normal-state paramagnetic excitations in Fe-based superconductors \cite{InosovPark10}. In addition, the low-temperature datasets at 3.5\,K and 15.5\,K also exhibit an unusual peak centered at $\hslash\omega_0=7$\,meV, which cannot be attributed to the conventional magnetic resonant mode because it persists almost unchanged above $T_{\rm c}$ and is located at an energy far above the anticipated onset of the particle-hole continuum ($2\Delta_1\approx4.3$\,meV, as follows from the analysis of our $\mu$SR data).

\makeatletter\renewcommand{\fnum@figure}[1]{\figurename~\thefigure.}\makeatother
\begin{figure}[t]
\centering
\includegraphics[width=\columnwidth]{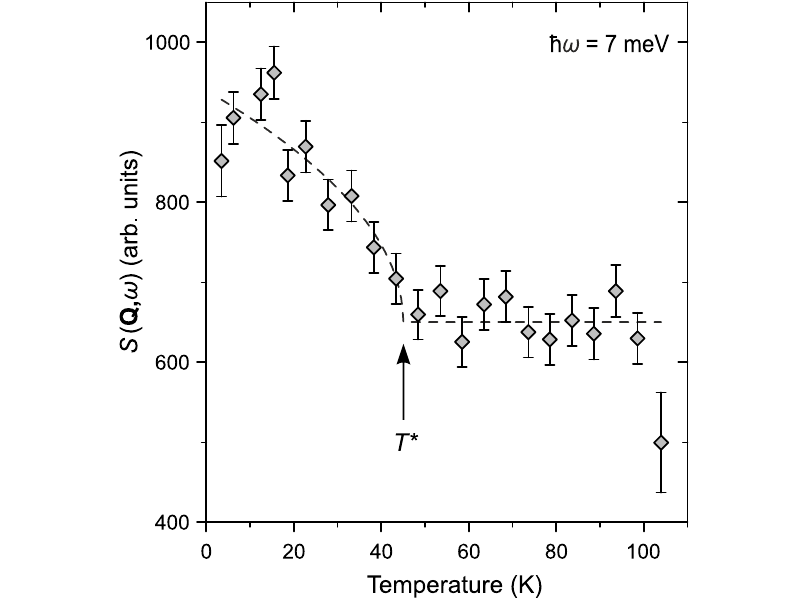}
\caption{Temperature dependence of the background-subtracted magnetic intensity at $\hslash\omega=7$\,meV, measured at $\textbf{Q}$=$\bigl(0 \half\bigr)$, demonstrating an anomaly related to the onset of additional intensity below $T^\ast=45$\,K.}
\label{fig:T_Dep_1}
\vspace{-0.2em}
\end{figure}
\makeatletter\renewcommand{\fnum@figure}[1]{\figurename~\thefigure~(color online).}\makeatother

In order to verify the origin of this new peak, we have investigated the temperature dependence of $S(\mathbf{Q},\omega)$ at $\omega=7$\,meV, which is presented in Fig.~\ref{fig:T_Dep_1}. The signal exhibits a clear anomaly around $T^\ast=45$\,K, associated with an onset of additional intensity at low temperatures. To the best of our knowledge, no phase transition in this temperature range has been announced in the literature for the Pt-doped 10-3-8 compound. Another weak anomaly in the $T$-dependence could be suspected at $T_{\rm c}$, where the intensity of the 7\,meV peak appears to be slightly suppressed, in agreement with Fig.~\ref{fig:SQw}. If this suppression is confirmed, it could indicate a possible competition of the phenomenon responsible for the 7\,meV peak with superconductivity.

The sum rule for magnetic neutron scattering,\vspace{-1pt}
\begin{equation}
\textstyle\int\!\!\!\int\!S(\mathbf{Q},\omega)\,{\rm d}\mathbf{Q}\,{\rm d}\omega\propto S(S+1),\vspace{-1pt}
\end{equation}
implies that the formation of the 7~meV peak below $T^\ast$ must be associated with a simultaneous depletion of spectral weight at lower energies, which can be associated with the opening of a pseudogap. Within this scenario, the peak would represent a precursor resonant mode or a pile-up of magnetic intensity around the pseudogap energy. This interpretation motivated us to perform NMR measurements on the same sample, which we will present below, substantiating the presence of a pseudogap with comparable magnitude also in the electronic structure.

\begin{figure}[t!]
\centering\includegraphics[width=\columnwidth]{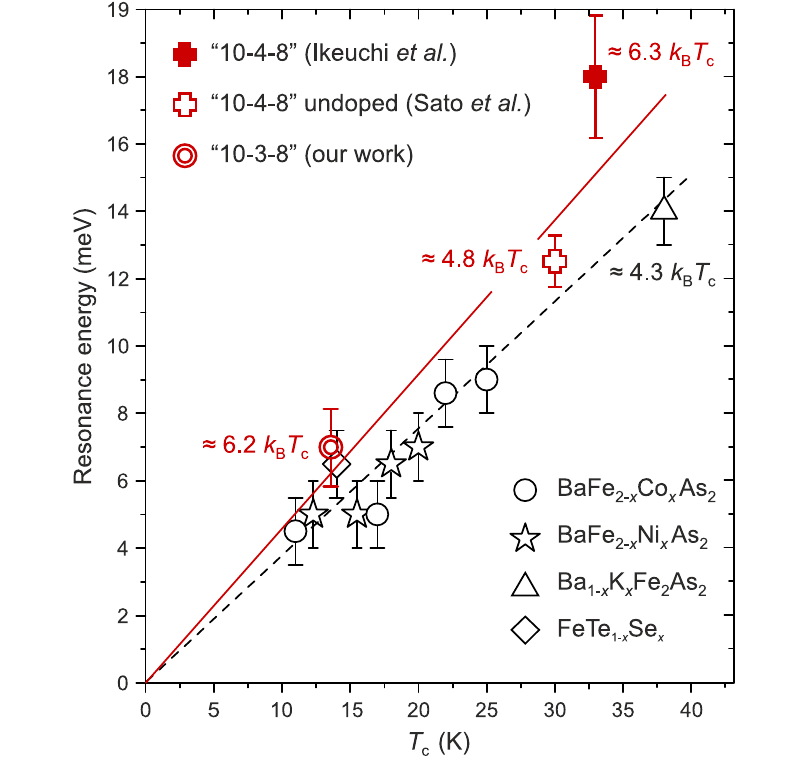}
\caption{Universal scaling of the resonant energy in 122-type iron-pnictide superconductors (dashed line), adapted from Ref.~\protect{\citenum{ParkInosov10}}. Energies of the low-temperature intensity maxima observed in 10-3-8 and 10-4-8 arsenides from Refs.~\protect{\citenum{Sato_Kawamata_2011, IkeuchiSato14}} and the present work are plotted for comparison to demonstrate that their $\hslash\omega_0/k_{\rm B}T_{\rm c}$ ratios are systematically higher than those expected for the magnetic resonant modes.\vspace{-3pt}}
\label{fig:Scaling2}
\end{figure}

It is illustrative to compare our INS results with those obtained previously on the related (CaFe$_{1-x}$Pt$_x$As)$_{10}$Pt$_4$As$_8$ compounds with higher values of $T_{\rm c}$ \cite{Sato_Kawamata_2011, IkeuchiSato14}. In the 10-4-8 sample with $T_{\rm c}\approx30$\,K investigated by Sato \textit{et al.}, a peak in $\chi''(\mathbf{Q}, \omega)$ was found at an energy of $\sim$\,12\,meV, which gained intensity upon cooling without exhibiting any clear anomaly at $T_{\rm c}$, as evidenced by Fig.\,5 in Ref.\,\citenum{Sato_Kawamata_2011}. This behavior is very similar to that observed in our 10-3-8 sample. A more recent report of the same group on another sample with $T_{\rm c}\approx33$\,K presents the observation of a similar peak around $\sim$\,18\,meV [Fig.\,2\,(b) in Ref.\,\citenum{IkeuchiSato14}]. We collected all these results in Fig.\,\ref{fig:Scaling2}, where we plot the peak energy vs. $T_{\rm c}$ together with the universal scaling relationship of the neutron resonant mode in 122-type iron pnictides \cite{ParkInosov10}, shown for comparison. Clearly, the energies of the peaks observed in all three iron-platinum arsenides fall systematically above the $\hslash\omega_{\rm res}\approx4.3k_{\rm B}T_{\rm c}$ line, which could indicate their common origin that is distinct from the usual superconducting resonant mode, but universal among the whole class of materials.

\vspace{-5pt}\subsection{NMR measurements}\vspace{-5pt}

\begin{figure}[b!]
\includegraphics[width=\columnwidth]{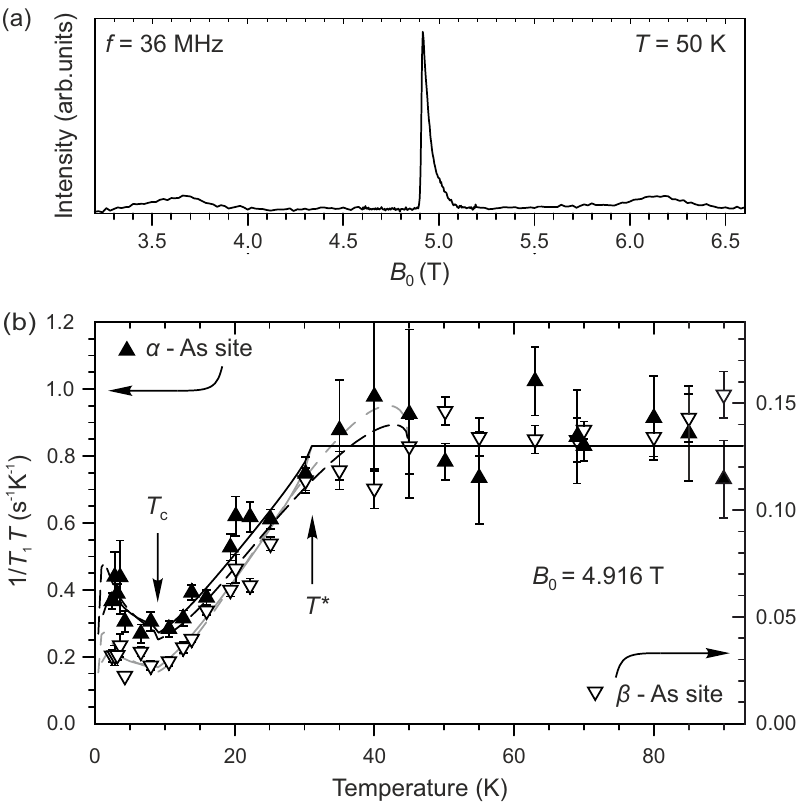}
\caption{(a)~$^{75}$As NMR spectrum at $T=50$\,K. (b)~Corresponding temperature dependent $1/T_1T$ data obtained under the assumption of two As sites: $\alpha$-As ($\blacktriangle$, black lines) and $\beta$-As ($\triangledown$, gray lines). The solid and dashed lines denote fits to the $s^\pm$-wave and $s$-wave pseudogap models, respectively, as described in the text. Note that both $T_{\rm c}$ and $T^{\ast}$ (shown with arrows) are reduced by the applied magnetic field, $B_0=4.916$\,T, from their zero-field values.\vspace{-3pt}}
\label{fig:NMR}
\end{figure}

Nuclear magnetic resonance is directly sensitive to changes in the electronic DOS in the vicinity of the Fermi level. The opening of small energy gaps can be precisely tracked by measuring the temperature dependence of the spin-lattice relaxation rate, $1/T_1$. Therefore, we performed both $^{75}$As NMR spectroscopy and $T_1$ relaxation experiments on two pieces of the same single crystal that was used for the INS study described above, with both samples showing consistent behavior. The $^{75}$As spectrum shown in Fig.~\ref{fig:NMR}\,(a) was taken with the external field applied orthogonally to the $ab$ plane, at frequency $f=36$\,MHz and temperature $T=50$\,K. Since $^{75}$As is a spin-$\threehalf$ nucleus with 3 possible transitions, a central line $\bigl(\half \leftrightarrow -\half\bigr)$ and two satellites $\bigl(\pm \threehalf \leftrightarrow \pm \half\bigr)$ are observed. The quadrupolar frequency is around $9.0~\mathrm{MHz}$, which is consistent with former work \cite{Zhou_Koutroulakis_2013}. Apart from a marginal broadening of the lines upon cooling, no significant changes of the spectrum with temperature could be detected. The small asymmetry of the central line is due to the slightly oblique mounting of the sample, which has no significant effect on the relaxation-rate measurements.

\makeatletter\renewcommand{\fnum@figure}[1]{\figurename~\thefigure.}\makeatother
\makeatletter\renewcommand{\fnum@figure}[1]{\figurename~\thefigure~(color online).}\makeatother

The $T_1$ relaxation experiments were carried out with the saturation recovery method at the maximum of the central line. The relaxation of the central line of a spin-$\threehalf$ nucleus can be described as in Ref.~\citenum{Suter_Mali_1998}. Since there are two inequivalent As sites in this compound (denoted henceforth as $\alpha$ and $\beta$) that generally can have different relaxation times, we have chosen the following fitting function for the recovery curves:\vspace{-3pt}
\begin{multline}\label{Eq:NMRfit}
m(t) = A - B \Bigg\lbrace C \bigl(0.9\,e^{-6t/T_1^{\alpha}} + 0.1\,e^{-t/T_1^{\alpha}}\bigr)\\
+ (1-C) \bigl(0.9\,e^{-6t/T_1^{\beta}} + 0.1\,e^{-t/T_1^{\beta}}\bigr) \Bigg\rbrace,\vspace{-8pt}
\end{multline}
where $A$ and $B$ are unassigned fitting parameters, $T_1^{\alpha}$ and $T_1^{\beta}$ are relaxation times for the $\alpha$- and $\beta$-As sites, respectively, and $C$ denotes the relative contribution of the $\alpha$-As site in the spectral point, which was assumed to be $T$-independent.

The $1/T_1T$ data extracted from the fit for both As sites are plotted in Fig.~\ref{fig:NMR}\,(b). At high temperatures, the constant $1/T_1T$ reveals Korringa behavior typical for a normal Fermi liquid. Below $T^\ast$, the value of $1/T_1T$ gradually drops to nearly a quarter of its original level, reaching a minimum around the superconducting transition temperature that is reduced here to $T_{\rm c}{\kern.7pt}(B_0) \approx 9$\,K by the applied magnetic field. At even lower temperatures, $1/T_1T$ is increasing slightly again. The over-proportional slowing-down of the relaxation can be interpreted as a loss of DOS around the Fermi level, usually associated with an energy gap. As its onset temperature lies well above $T_{\rm c}$, it has to originate from a pseudogap that persists already in the normal state under the assumption that the value of $T^\ast$ is somewhat suppressed by magnetic field as compared to its zero-field value extracted from the INS data.

More specifically, our $1/T_1T$ data can be equally reconciled with two pseudogap models, which are shown in Fig.\,\ref{fig:NMR}\,(b) by the solid and dashed fitting lines. The first model (solid lines) assumes an $s^\pm$ character of the pseudogap and a reduced $T^\ast=31$\,K as compared to the zero-field result from INS. The second model (dashed lines) assumes an $s$-wave pseudogap with $T^\ast=45$\,K that is unchanged with respect to the zero-field value. While both models agree equally well with the data, the appearance of a precursor resonance-like peak in the INS signal would be easier to reconcile with the sign-changing $s^\pm$-type symmetry of the pseudogap. In both fitting models for the relaxation rate, we have included the pseudogap, which was treated similarly to a superconducting BCS gap \cite{Williams_Tallon_1997}, along with the second (smaller) superconducting gap revealed in our $\mu$SR experiment, so that
\begin{equation}
\frac{1}{T_{1}T}\propto \frac{2}{k_\mathrm{B}T}\!\int\limits_{0}^{\infty} \bigl[N(\omega)^{2} + M(\omega)^{2}\bigr]\,f(\omega) [1-f(\omega)]\,{\rm d}\omega.
\end{equation}
Here $N(\omega)$ and $M(\omega)$ are the energy-dependent normal and anomalous DOS distributions in the vicinity of the Fermi level, and $f(\omega)$ is the Fermi function. $N(\omega)$ and $M(\omega)$ can be described with a two-gap model. The pseudogap develops at $T^\ast=31$\,K in the $s^\pm$ pseudogap model (or $T^\ast=45$\,K in the $s$-wave model) and follows an order-parameter-like behavior given by Eq.~\ref{fig:T_Dep_1}. Its magnitude amounts to $\Delta_{\mathrm{PG}}=3.9$\,meV ($s^\pm$-wave) or 4.8\,meV ($s$-wave) in the low-temperature limit. The assumption of an $s$-wave symmetry of the pseudogap results in a broad Hebel-Slichter peak emerging below $T^{*}$ because of the sharp onset of the pseudogap implied by our model, whereas no such peak appears in the $s^\pm$ case. The DOS fraction of the ``pseudogapped'' band is $52\%$ for $\alpha$-As and $66\%$ for $\beta$-As sites in the $s^\pm$-wave model ($47\%$ and $60\%$, respectively, for the $s$-wave model).

A second gap with an $s$-wave symmetry develops at the superconducting transition temperature, $T_\mathrm{c}(B_0)=9$\,K. It is very small, with $\Delta_{\rm SC}(B_0)=0.136$\,meV or $2\Delta_{\rm SC}/k_\mathrm{B}T_\mathrm{c}(B_0)=0.35$, which is in reasonable agreement with the smaller gap found in our $\mu$SR experiments. Remarkably, no sharp anomaly in the relaxation rate is observed at $T_{\rm c}$, unlike in most other Fe-based superconductors \cite{LongWeiQiang13}. Independently of the assumed order-parameter symmetry, the inclusion of the second (larger) superconducting gap in the model does not lead to any considerable improvement of the fit. This indicates that the larger gap causes no significant reduction in the DOS below the superconducting transition or, in other words, that the corresponding electron bands are fully gapped already above $T_{\rm c}$. At the same time, the opening of the smaller superconducting gap can be clearly seen in the data, as it leads to the emergence of a Hebel-Slichter peak at very low temperatures ($\sim 2$\,K) in good agreement with our model.

\vspace{-5pt}\section{Discussion}\vspace{-5pt}

We note that signatures of a pseudogap formation were reported earlier from NMR measurements on several other Fe-based superconductors, most notably in oxypnictides with the 1111-type structure \cite{NakaiIshida08, GrafePaar08, AhilanNing08} and alkali-metal iron selenides \cite{MaJi11}, where it has been usually attributed to the presence of magnetic correlations \cite{GrafeLang08, IshidaNakai_book}. There are several important differences, however, between these observations and the pseudogap-like behavior of the Pt-doped 10-3-8 superconductor presented here. First, the monotonic increase of $1/T_1T$ in oxypnictides is usually observed over a much broader temperature range and persists up to room temperature, without any sharp onset that could be linked to an anomaly in the spin-fluctuations spectrum to verify its magnetic origin. Second, the relaxation rate there still shows sharp anomalies at $T_{\rm c}$, indicating that the pseudogap possibly resides at a different part of the Fermi surface or is only partially open immediately above the superconducting transition. Finally, our INS measurements reveal no dramatic change in the spin-fluctuation spectrum across $T^\ast$ apart from the formation of a small peak at $\hslash\omega_0$ with too little spectral weight to account for the breakdown of the Korringa behavior. This peak can be rather seen as a consequence of the pseudogap formation, representing a magnetic spectral-weight redistribution by analogy with the spin resonance emerging below $2\Delta$ in most other iron-based superconductors within the superconducting state. Our estimate of the pseudogap magnitude from NMR is fully consistent with this interpretation, as $2\Delta_{\rm PG}=7.8$\,meV ($s^\pm$-wave model) lies immediately above the INS peak seen in Fig.\,\ref{fig:SQw}. Similarly, $\mu$SR revealed no significant increase in the muon relaxation rate at low temperatures that could signal critical slowing down of spin fluctuations in the proximity to a magnetic instability, as actually expected for an optimally doped sample sufficiently remote from the AFM dome in the phase diagram. This renders the magnetic origin of the pseudogap in our 10-3-8 sample unlikely.

An alternative scenario of ``preformed pairs'', which was first put forward (among others) to explain the pseudogap in high-$T_{\rm c}$ cuprates \cite{EmeryKivelson95, YangRameau08, TimuskStatt99, HashimotoVishik14}, or a Berezinskii-Kosterlitz-Thouless (BKT) kind of transition expected in low-dimensional systems \cite{MadanKurosawa14}, seems to be much more consistent with the presented results. In these scenarios, the formation of phase-incoherent Cooper pairs occurs already above $T_{\rm c}$, leading to a depletion of the DOS at the Fermi level on an energy scale comparable with the superconducting gap. In our case, this condition is fulfilled, because the ratio $\Delta_{\rm PG}/k_{\rm B}T^\ast\approx1.46$ ($s^\pm$-wave model) is rather close to the coupling constant of 1.76 predicted by the BCS theory. Moreover, our estimate for $T^\ast=45$\,K is not too far from the maximal $T_{\rm c}$ of $\kern.5pt\sim\kern.8pt$35\,K among iron-platinum arsenides \cite{Sturzer_Derondeau_2014}. The preformed-pairing mechanism can also naturally explain the absence of an additional anomaly from the superconducting gap opening at $T_{\rm c}$ in our NMR data, as the common origin of the two gaps implies that the large gap is fully open already above $T_{\rm c}$. Conversely, the smaller gap, which could either reside on one of the Fe\,$d$ bands near the $\Gamma$ point \cite{EvtushinskyInosov09} or be induced by the proximity effect on the Pt\,$d$ Fermi surfaces \cite{Berlijn_2014}, forms only below $T_{\rm c}$, resulting in a Hebel-Slichter anomaly around 2\,K. Coexistence of such a peak with a resonance-like feature in the INS data is highly unusual among Fe-pnictides and could speak in favor of a difference in symmetry between the larger and smaller gaps. In contrast to the $1/T_1T$ relaxation-rate measurements probing the DOS at the Fermi level, ac susceptibility [Fig.~\ref{fig:magnetization}\,(b)] and $\mu$SR [Fig.~\ref{fig:muSR}] experiments are sensitive to the superfluid density, i.e. only to the phase-coherent fraction of Cooper pairs that underwent Bose-Einstein condensation below $T_{\rm c}$. This explains why both measurements are insensitive to the pseudogap opening and display conventional behavior.

In summary, our results point toward possible preformed Cooper pairing in platinum-iron arsenides, leading to a pseudogap formation below $T^\ast$ on parts of the Fermi surface responsible for $\sim\kern.7pt$50--60\% of the DOS, which becomes fully open upon reaching $T_{\rm c}$. Below $T_{\rm c}$, the pairs gain phase coherence and form a superconducting condensate, as can be seen in the diamagnetic response and in the sharp increase of the superfluid density. Simultaneously, a second (smaller) superconducting gap is induced on the remaining parts of the Fermi surface. Quantitative assessment of all the mentioned gaps and transition temperatures has been presented. The spin-fluctuation spectrum responds to the pseudogap formation by a partial redistribution of the low-energy spectral weight, leading to a pile-up of magnetic intensity around 7~meV that resembles a precursor magnetic resonant mode.

\vspace{-5pt}\section*{Acknowledgments}\vspace{-5pt}

We would like to thank J.~White for his help with the crystal alignment at the Morpheus diffractometer (PSI, Switzerland). D.\,S.\,I. appreciated stimulating conversations with A.~Chubukov, S.-L.~Drechsler, and D.~Efremov. This project was financially supported by the German Research Foundation (DFG) within the Grant No.~SA~\mbox{2426/1-1}, priority program SPP~1458/2 (Grants No.~IN209/1-2 and KL1086/10-2), and the Graduiertenkolleg GRK~1621 at the TU Dresden, as well as by the Basic Science Research Program (NRF-2013R1A1A2009778) and the Leading Foreign Research Institute Recruitment Program (Grant No. 2012K1A4A3053565 through NRF).

\end{document}